\documentclass[12pt]{article}

\usepackage[pdftex]{graphicx}
\usepackage{amssymb}

\newcommand{\beq}{\begin{equation}}
\newcommand{\eeq}{\end{equation}}
\newcommand{\bea}{\begin{eqnarray}}
\newcommand{\eea}{\end{eqnarray}}

\newcommand{\g}{\gamma}

\renewcommand{\b}{\beta}

\begin{document}

\begin{center}
\vspace{48pt}
{ \Large \bf  Pseudo-topological transitions in 2D
 gravity models coupled to massless scalar fields }

\vspace{40pt}

{\sl J.\ Ambj\o rn}$\,^{a,c}$,
{\sl A.\ T. G\"{o}rlich}$\,^{a,b,d},$
{\sl J.\ Jurkiewicz}$\,^{b,e}$
and {\sl H.-G.\ Zhang}$\,^{b,f}$

\vspace{24pt}

{\small

$^a$~The Niels Bohr Institute, Copenhagen University\\
Blegdamsvej 17, DK-2100 Copenhagen \O , Denmark.

\vspace{10pt}

$^b$~Mark Kac Complex Systems Research Centre,\\
Marian Smoluchowski Institute of Physics, Jagellonian University,\\ 
Reymonta 4, PL 30-059 Krakow, Poland.
}

\end{center}


\vspace{36pt}

\begin{center}
{\bf Abstract}
\end{center}

We study the geometries generated by 
two-dimensional causal dynamical triangulations (CDT) coupled 
to $d$  massless scalar fields. 
Using methods similar  to those used to study four-dimensional CDT 
we show that there exists a $c=1$ ``barrier'', analogous to the $c=1$
barrier encountered in non-critical string theory, only 
the CDT transition is easier 
to be detected numerically. For $d\leq 1$ we observe 
time-translation invariance and geometries entirely governed by 
quantum fluctuations around the uniform toroidal topology put 
in by hand. For $d>1$  the effective average geometry is no 
longer toroidal but  ``semiclassical''  and  spherical with Hausdorff 
dimension $d_H = 3$. In the $d>1$ sector we study the
time dependence of the semiclassical 
spatial volume distribution and show that the observed 
behavior is described an effective  mini-superspace
action analogous to the actions found  in the 
de Sitter phase of three- and four-dimensional  
pure CDT simulations 
and in the three-dimensional CDT-like Ho\v rava-Lifshitz models.  

\vspace{12pt}
\vfill

{\footnotesize
\noindent
$^c$~{email: ambjorn@nbi.dk}\\
$^d$~{email: goerlich@nbi.dk, atg@th.if.uj.edu.pl}\\
$^e$~{email: jerzy.jurkiewicz@uj.edu.pl}\\
$^f$~{email:  zhang@th.if.uj.edu.pl}\\

}


\newpage

\section{Introduction}\label{intro}

The formalism of dynamical triangulations (DT) \cite{old} was introduced
to provide a lattice regularization of Polyakov's theory 
of non-critical strings \cite{polyakov} and 2d quantum gravity, and later of 
both 3d and 4d quantum gravity \cite{old1}. It was successful
in the 2d case, but no interesting continuum limit has so far
been  found in the pure gravity sector of
the higher dimensional lattice quantum gravity theories \cite{firstorder}.
The problem was that two unphysical sectors, the so-called ``crumpled phase''
where geometries have a Hausdorff dimension $d_H = \infty$, and the 
branched polymer (BP) phase, where the Hausdorff dimension of the geometries
was $d_H = 2$, where separated by a first order phase transition\footnote{
Attempts to obtain different Hausdorff dimensions by adding matter 
or using different weight factors for the path integral led to 
the possibility of a new phase, the ``crinkled phase'' \cite{crinkled},
which seems to have Hausdorff dimension close to four.}.

Causal Dynamical triangulations (CDT) \cite{al1,ajl3d,ajl4d} 
were introduced to cure 
the problems  encountered in the three- and four-dimensional
Euclidean DT quantum gravity theories. The idea was to insist on
a (proper time) foliation\footnote{The idea 
of a time foliation as important part of a theory of quantum gravity  
has also been advocated  recently by Ho\v rava  \cite{horava1}. The 
Ho\v rava-Lifshitz theory of gravity might be closely related to 
the CDT theory of gravity. One observes the same spectral dimension 
in the four-dimensional theories \cite{spectral1,spectral2}, and the  same
kind of phase diagram \cite{al-samo}. Also in three dimensions there is 
a  close connection \cite{newD3}.}, which would curb the formation of 
some of the ``pathological'' DT configurations \cite{problems}, 
and indeed some success has been obtained. 
A semiclassical regime of bare coupling 
constants has been located (the so-called de Sitter phase) \cite{semiclassical},
quantum universes of the size of 10-20 Planck units have been 
studied \cite{planck}, and a second order phase transition line, where 
a continuum UV limit might exist, has been located \cite{secondorder}.

Except for a few features of 3d CDT \cite{3dCDT}, all the studies 
of the higher dimensional CDT theories have used computer simulations
and the precise relation of the CDT theories to the old DT theories
is not clear\footnote{A recent attempt \cite{laiho} to revive the so-called 
crinkled phase of DT may hint a connection. One observes a 
behavior of the spectral dimension which is almost identical to
the one observed in CDT.}. In this respect two-dimensional CDT
is different. This theory can be solved analytically and its
relation to the two-dimensional Euclidean quantum gravity, as 
solved by the DT model, is also clear. The CDT theory can be 
viewed as an effective theory, where baby-universes created in the 
DT-theory have been integrated out \cite{ack}. As a result the  
Hausdorff dimension of the ensemble of 2d CDT geometries is 
equal to the canonical dimension, $d_H=2$, while the Hausdorff
dimension of the ensemble of DT geometries is four\footnote{It is 
possible to introduce an additional coupling constant related
to the creation of baby universes and follow the detailed transition
from CDT geometries to DT geometries \cite{generalizedCDT}.} 
\cite{transfer,aw,ajw,bowick,nishimura}. 

One of the amazing features 
of the DT theory of two-dimensional quantum gravity is that 
it can be solved analytically for any minimal rational conformal field
theory coupled to geometry, as long as the central charge 
$c$ of the conformal theory is less than or equal to 1, 
and one finds agreement  with the corresponding continuum results 
from quantum Liouville theory \cite{KPZ}. For $c >1$ the analytical 
solutions become unphysical (complex critical exponents) and 
it is believed that the strong interaction between matter 
and geometries forces the geometries to degenerate into 
branched polymers, an observation going all the 
way back to the study of random surfaces on hyper-cubic lattices 
and random surfaces using DT \cite{durhuus,ad}.   

While it is straightforward
to couple matter to the 2d CDT theory \cite{aal,aalp}, 
one can unfortunately not (yet) solve the theory analytically. 
However, numerically one can still investigate the theory.
One finds that for the Ising model and the three-state Potts model
coupled to CDT, which have $c=1/2$ and $c=4/5$,
the critical exponents of matter remain equal to the flat space
exponents. This is in sharp contrast to the situation when using
the 2d DT regularization, 
where the critical exponents change from their flat space
values to their so-called KPZ values \cite{KPZ}. Since the ensemble
of CDT geometries can be derived from the DT ensemble of geometries 
by integrating out baby universes, it is natural to associate the 
KPZ exponents with baby universe creation, an intuition which
has been made quite concrete for the entropy exponent $\g$ 
\cite{jain,ajt,adj}, and an observation made before the invention of 
the CDT formalism. Thus it is natural to conjecture that CDT coupled
to conformal field theories with $c \leq 1$ can still be viewed
as  effective DT theories where baby universes have been integrated out,
although no analytic proof exists presently. The interesting question
which we want to address in this article is what happens in CDT 
when one crosses the $c=1$ barrier. As mentioned above, the KPZ 
exponents become complex, signaling a breakdown of the 
concept of a two-dimensional surface on which the conformal 
matter lives into branched polymers. The formation of 
local spikes \cite{spikes} when $c >1$ is believed to be the 
explanation of this world sheet instability.

In the CDT case the
geometry cannot degenerate into BP by construction. 
Is there a $c=1$ barrier at all?
In \cite{aal} a hint of a different behavior for $c>1$ was seen:
for 8 Ising spins (i.e. $c=4$) coupled to CDT it 
was observed that the average geometry was very different from 
the $c<1$ geometry. In the following we will provide evidence
that the observation made in \cite{aal} is not accidental,  
that the geometry undergoes a phase transition for $c>1$, and 
that this transition is easier to observe 
from a numerical point of view than the ``old'' DT transition. The transition
is not to a BP phase, but to a phase of ``spherical, semiclassical'' 
geometry, which resembles somewhat the geometry encountered in higher 
dimensional CDT.

\section{ Choice of matter and numerical setup}\label{Setup}

\begin{figure}[t]
\centerline{\scalebox{0.5}{\includegraphics{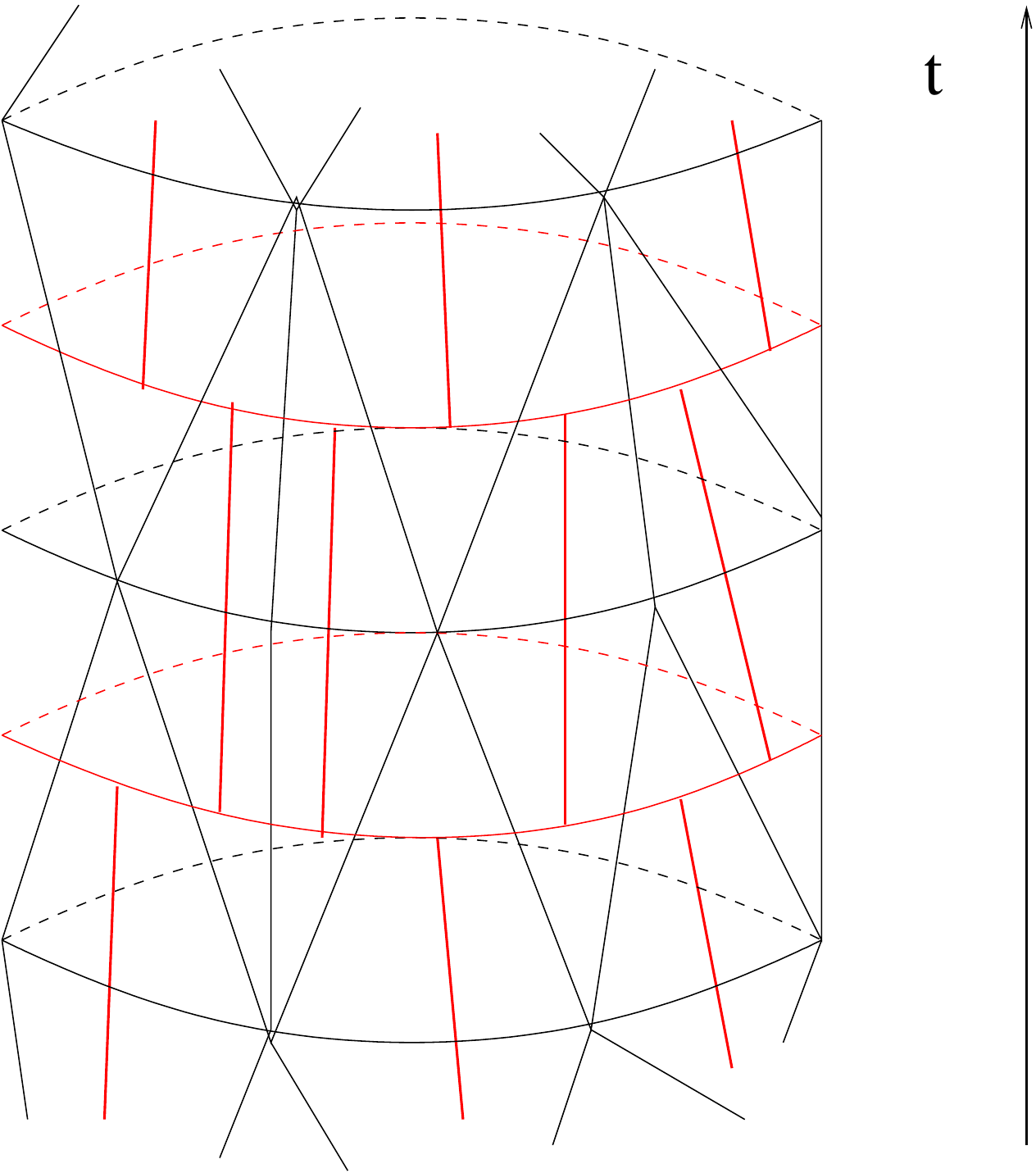}}}
\caption{ Direct and dual lattices of Causal Dynamical Triangulations in 2D.}
\label{duality}
\end{figure}


In order to study the effect of conformal matter on the CDT geometry
we have chosen to use massless scalar fields rather than 
the Ising spins used in \cite{aal}. Using Ising spins is more
demanding from a numerical point of view. The behavior 
of the Ising system is governed by the strength of the spin-spin
coupling\footnote{To be more precise, denoting the spin 
coupling $J$, we have $\b =J/kT$, where $T$ is the temperature
and $k$ Boltzmann's constant.} $\b$. For large $\b$ the system 
is magnetized and effectively decouples from the geometry. Similarly,
for small $\b$ the spins flip uncorrelated, again largely independent
of the geometry. At a certain critical $\b=\b_c$ there is a transition 
between the magnetized and demagnetized phase, and only at this 
point there is a significant coupling between geometry and matter,
corresponding to a $c=1/2$ conformal field theory coupled to geometry. 
However, we first have to locate $\b_c$ numerically, and contrary to 
the situation for a fixed geometry, $\b_c$ will depend on the number of 
Ising spins we use. Massless Gaussian fields are automatically critical 
and we have no such fine-tuning problems.

The geometric setup is the following: we want to have a time 
foliation of our 2d geometries with a given proper time $T$. 
We assume that  time has been Wick-rotated such that our 2d 
space-time has Euclidean signature, and finally we assume
that time is periodic. In the CDT
discretization we thus discretize the proper time in $T$ discrete
units of length  $a$, $a$ being the link length of the {\it equilateral}
triangles we will use. $a$ will play no role in the sense that 
it can be absorbed in coupling constants with dimension and we choose
$a=1$ in the following. Thus the discretized time takes the values
$t=0,\ldots,T-1$. 
The spatial slice at time $t$ is assumed to have the topology of $S^1$
and in the CDT regularization it consists of $L(t)$ links. 
The space-time slab between $t$ and $t+1$ will then be formed 
by a triangulation of $L(t)$ triangles which have two vertices at $t$ 
and one vertex at $t+1$, and $L(t+1)$ triangles with two vertices
at $t+1$ and one vertex at $t$, these $V(t) = L(t)+L(t+1)$ 
triangles glued together such that the topology of the slab 
is $S^1 \times [0,1]$. Since time is taken periodic the 
global topology of our 2d Euclidean space-time will be that
of the torus, $T^2= S^1\times S^1$ and the total space-time 
volume is proportional to the total number of triangles
\begin{equation}\label{N}
N = \sum_{t=0}^{T-1} V(t).
\end{equation}

A natural way to couple matter to geometry in CDT is to 
place the matter fields  at the center
of the triangles. Since we are using equilateral triangles
the Gaussian  action-term associated with such matter assignment
for a given triangulation ${\cal T}$is trivial: 
\begin{equation}\label{Gauss}
S_{Gauss} ({\cal T},x^\mu) = \sum_{\langle i,j \rangle} \sum_{\mu =1}^d 
(x_i^\mu -x_j^\mu)^2.
\end{equation} 
Here the summation is over $d$ Gaussian fields, and over neighboring triangles
$i,j$. A configuration in the path integral is characterized by 
a geometry, i.e.\ a triangulation ${\cal T}$, and a field configuration
assigned to that geometry, i.e.\ $d\times N$ values $x_i^\mu$. The 
probability amplitude, or 
(since we work with Euclidean signature) the Boltzmann weight,
of the configuration is then 
 \begin{equation}\label{probability}
P({\cal T}) \propto e^{-S({\cal T})},\quad S({\cal T}) = 
S_{Gauss}({\cal T},x^\mu) + 
\Lambda N({\cal T}) +\epsilon(N({\cal T})-\bar{N})^2.
\end{equation}
Here $\Lambda$ denotes a cosmological term and term 
$\epsilon(N-\bar{N})^2$ is added to insure that the 
space-time volume fluctuates around a prescribed value $\bar{N}$.

Finally the path integral (the state sum or partition function 
in the statistical mechanics interpretation) is formally
\begin{equation}\label{pathintegral}
{\cal Z} = \sum_{\cal T} \frac{1}{C_T}\;
\int \prod_{i,\mu} dx_i^\mu~ e^{-S({\cal T})},
\end{equation}
where $C_T$ is a symmetry factor of the graph ${\cal T}$, the 
order of the so-called automorphism group of ${\cal T}$.
There is no need of a coupling constant in front
of the Gaussian action (as already mentioned) since such a coupling can be 
viewed as a rescaling  $ c\; x_i \to y_i$  which in the path integral can 
be compensated by a change of $\Lambda \to \Lambda' = \Lambda - d\log c$.
The path integral (\ref{pathintegral}) contains a zero mode
of the field, namely the constant field,  which 
has to be fixed in order for the integral to be well defined.

Finally, since we have placed the matter fields at the center
of the triangles, it is natural to use the graph dual to the 
triangulation, i.e. a $\phi^3$ graph. Triangles are then represented
as vertices and what we in such a $\phi^3$ graphs will 
call a spatial line, corresponds to half-integer times
in the triangulation picture, being of length $V(t)$ (see
Fig.\ \ref{duality}). The dual graph, which we also denote ${\cal T}$,
thus has $N({\cal T})$ vertices $i$, and each vertex has two ``space-like''
links (horizontal links) and one ``time-like'' (vertical) link pointing either 
up or down. This description of CDT using dual graphs was first 
used in \cite{charlotte}. 

We can use Monte Carlo simulations to calculate the 
expectation values of certain ``observables'' in 
the statistical ensemble defined by eq.\ (\ref{pathintegral}).
In order to use Monte Carlo simulations we have to be 
able to move ergodically between the configurations. 
This means we must be able to change geometry  and 
field configurations in such a way that successive changes can
bring us to any configuration. If changes made on the configurations
satisfy the so-called detailed balance condition successive updatings
will under quite general conditions lead to the correct probability 
distribution, independent of the starting configuration \cite{barkama}.

The geometric update on the $\phi^3$ graph is performed using two so-called 
moves: the move "add", which 
adds a new vertical link (and the two  corresponding vertices) 
and the inverse  move "del", where the vertical link (and the two 
corresponding vertices) is deleted. The two moves satisfy
the detailed balance condition and the geometric constraints, 
which do not permit to break  the system into disjoint parts.
As already mentioned the scalar fields $x_i^{\mu},~\mu=1,\dots d$ 
are localized at the vertices.  
They are assumed periodic both in space and time. 

In numerical simulations the field values enter the detailed balance 
condition. Geometric moves are supplemented by the local update of 
the matter fields,
performed after a sweep composed of $N$ "add" and "del" moves, 
performed with the same a priori probability. 
The matter fields are updated using the heat bath algorithm.
In the discussion below we shall be concentrated only on the distribution 
$V(t)$ as a function of a discrete time $t$. We shall integrate over all
values of the fields. The distributions are independent 
of the zero modes of the scalar fields.

\section{Numerical results}\label{Results}

\begin{figure}[t]
{\scalebox{0.55}{\includegraphics{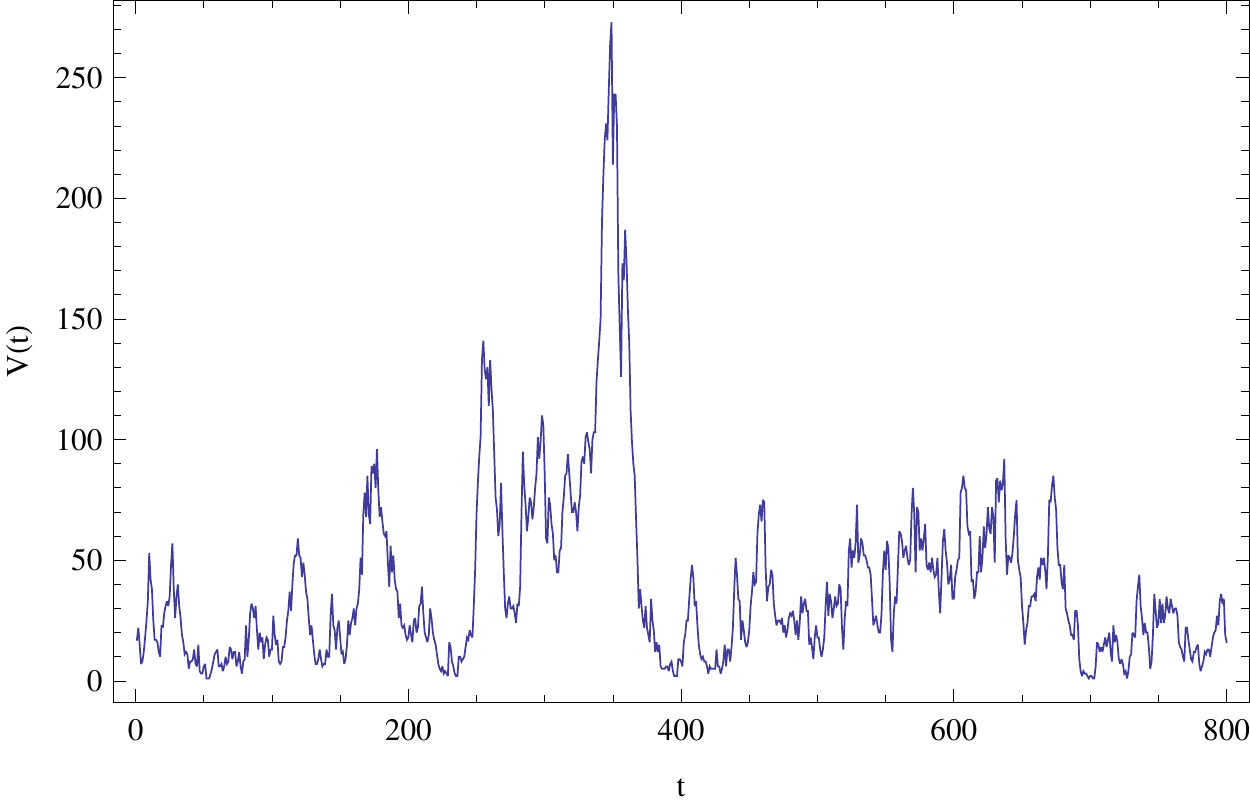}}}
{\scalebox{0.55}{\includegraphics{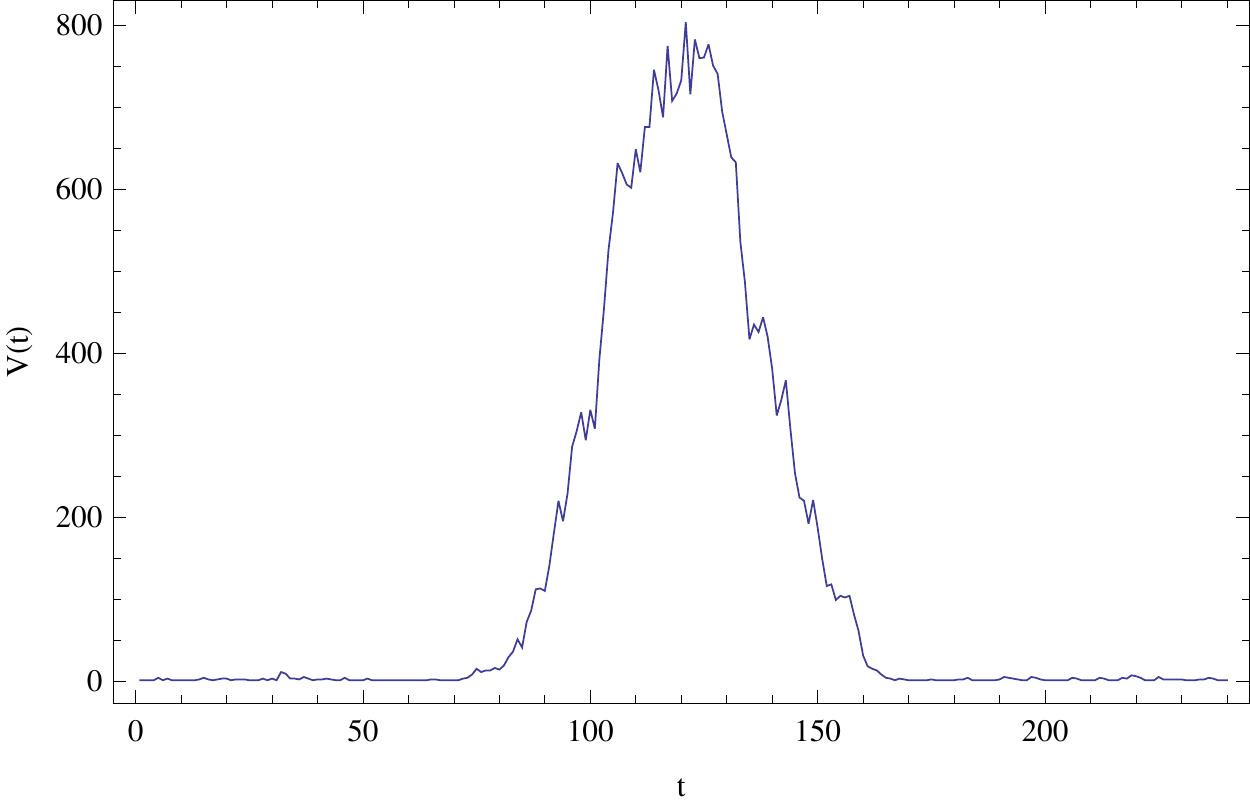}}}
\caption{Examples of configurations with one and four scalar fields coupled to CDT geometry.  The plots are  for $N = 64000$ and respectively $T = 800$ and $T = 240$.}
\label{config4}
\end{figure}

\subsection{The semiclassical configurations for \boldmath{$d >1$}}

If a class of semiclassical field configurations dominates the path integral,
expanding around any such configuration will break translational invariance
unless it is the constant field configuration. Usually the action 
is  translationally invariant and the zero-modes associated 
with this invariance should be treated as collective coordinates which 
will restore the invariance of the partition function when expanded 
around semiclassical solutions.  Rather surprisingly, we will 
encounter such a situation when $d > 1$. 
 
When $d >1$ and we just look at a typical geometric configuration,
as it presents itself in the path integral, we observe a ``blob''
if we plot $V(t)$ as a function of the proper time $t$. By a ``blob''
we mean that almost all the spatial volume is located in a region
of finite extension, say $\Delta T$, which is independent of $T$
if only $T$ is sufficiently large. $\Delta T$ will depend on $N$
and grow with $N$ as described below, but for a fixed $N$ $\Delta T$
is, up to fluctuations, the same for all {\it typical} configurations,
i.e.\ configurations we pick randomly from the computer\footnote{Of course
{\it all} configurations are present in the path integral, also configurations
which spread out over the whole range $T$. However, when $N$ is large
the probability they will be picked by a random selection will be very small.}.
Outside an interval of size $\Delta T$ $V(t)$ is basically zero, only 
we do not allow,{\it by construction}, the universe to shrink to zero spatial 
size. Thus we have outside $\Delta T$ a {\it stalk} of cut-off size,
superimposed with small fluctuations. To illustrate this behavior we 
show plots representing typical spatial 
volume distributions $V(t)$ versus time for 
$d=1$ and $d=4$ in Fig.\ \ref{config4}. For $d=1$ we do not observe 
the blob structure.

\begin{figure}[t]
{\scalebox{0.7}{\includegraphics{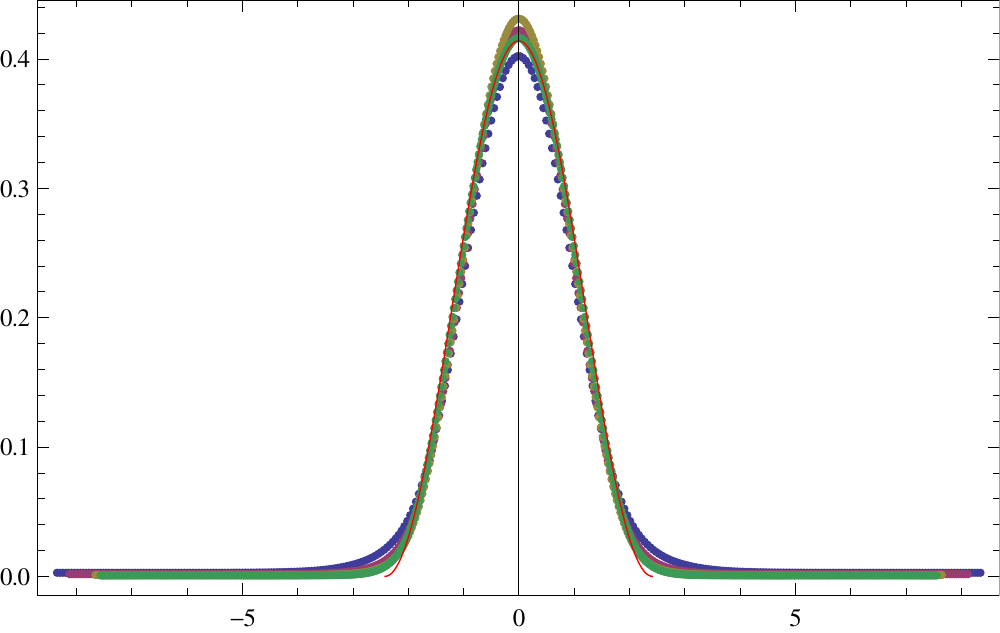}}}
{\scalebox{0.7}{\includegraphics{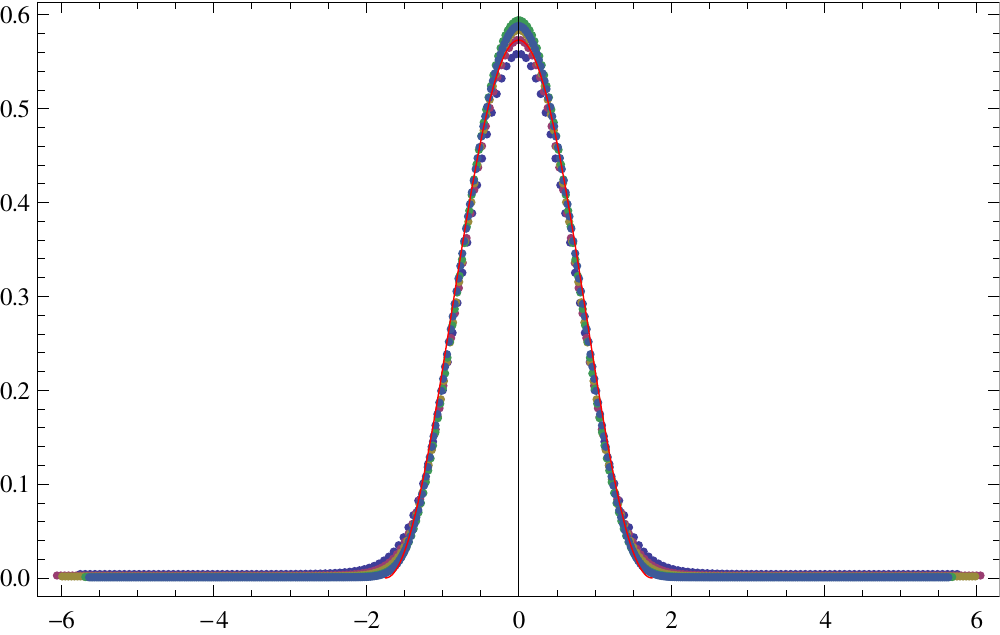}}}
{\scalebox{0.7}{\includegraphics{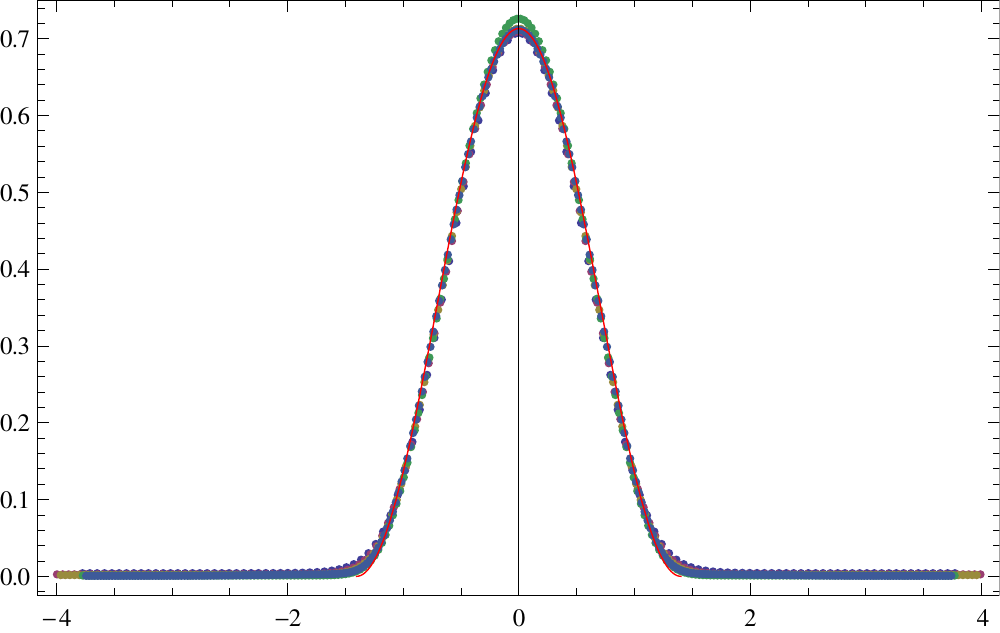}}}
{\scalebox{0.74}{\includegraphics{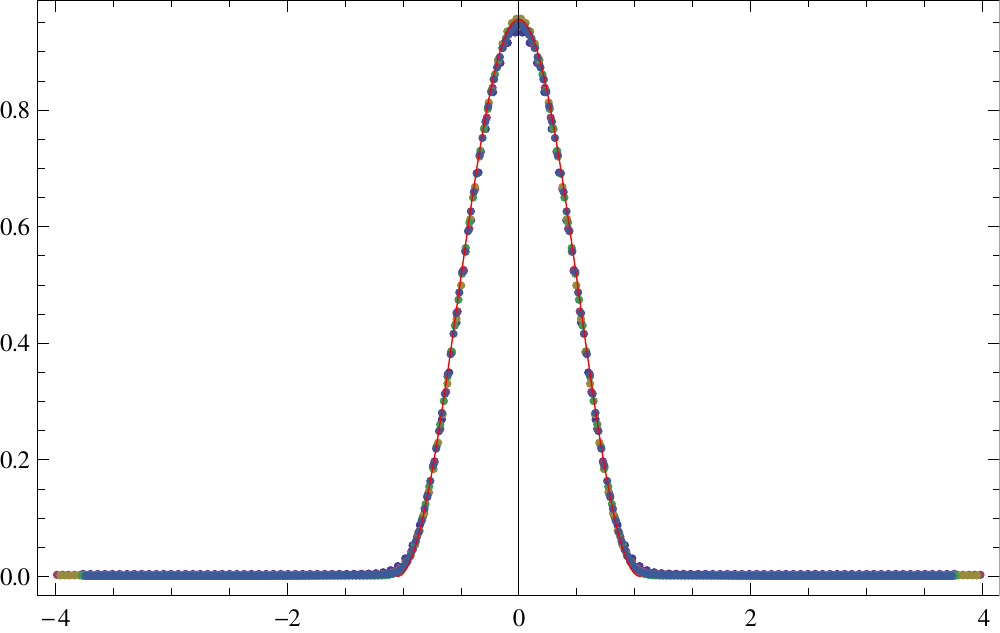}}}
\caption{
Universality of the rescaled volume distributions for 
various number of scalar fields. The top figures  represent 
$d=2$ and $d=3$ for $N = 8000, ~16000,~32000,~64000,~128000$. 
The time $T$ used depended on volume and was
respectively $T= 320,~ 400,~ 480,~ 600,~ 750, ~1000$ for $d=2$ 
and $T = 150, ~180,~240, ~300,~ 360, ~450$ for $d=3$. The bottom
figures correspond $d=4$ and $d=6$ for the same range of spatial volumes. 
The time was in both cases
$T = 100, 120,~160,~ 200,~ 240, ~300$.}
\label{scaling}
\end{figure}

The ``center of mass'' of the blob will perform a 
random walk as a function of computer time, due to translational time 
invariance of the action. Thus an average of $V(t)$ will just 
result in a uniform distribution. Clearly this will not capture the 
true nature of the configurations for $d>1$, and we have in fact
to ``undo'' the collective mode integration mentioned above associated 
with an expansion around a semiclassical solution. For 
each configuration we thus fix the position of the ``center of mass''
of the spatial volume distribution to be at $t=0$ and define time
to be symmetric around $t=0$.  We then 
average the distribution over many statistically independent configurations, 
obtaining in this way what we call a ``semiclassical distribution'', 
in the sense that configurations with the ``blob'' spatial volume profiles
are  the dominant ones in the path integral. 
  
\begin{figure}[t]
\centerline{\scalebox{0.8}{\includegraphics{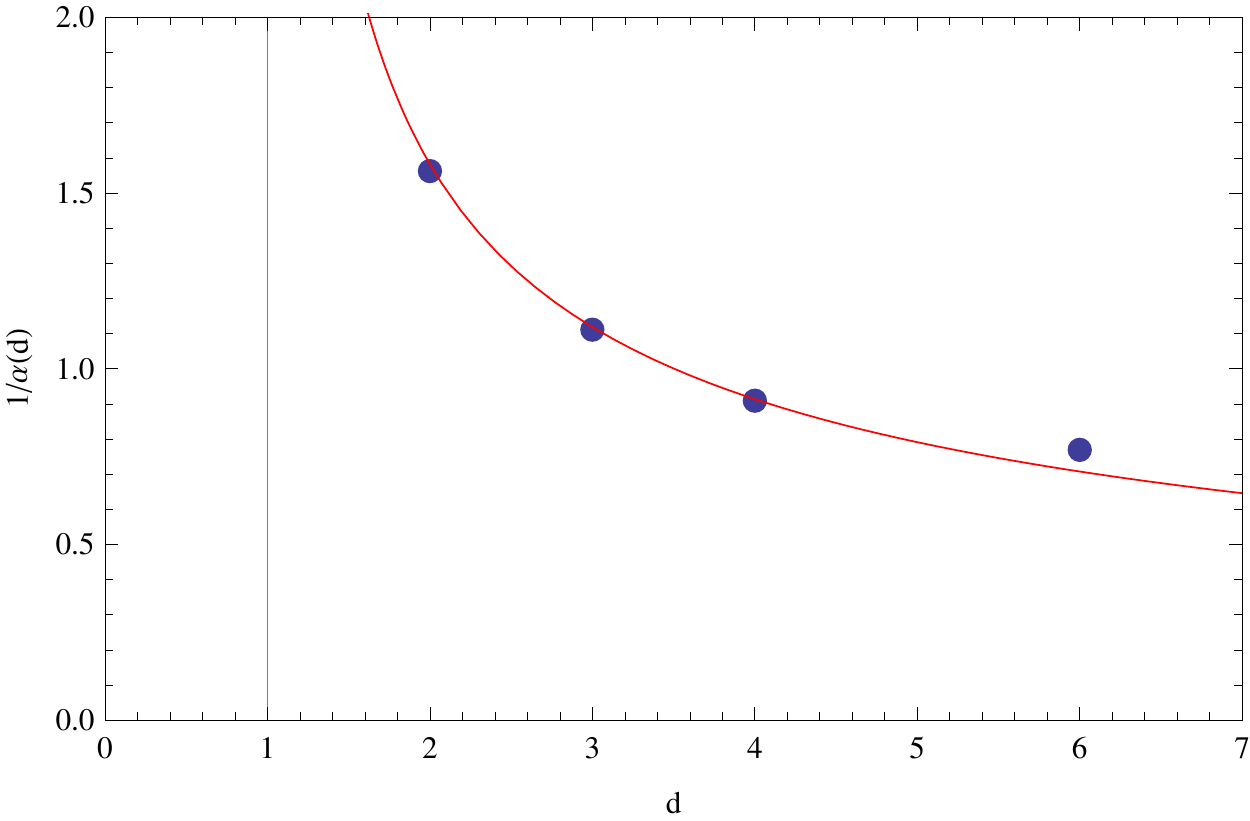}}}
\caption{The dependence of  $1/\alpha$ on $d$ is shown. 
The line is ${\rm const.}/\sqrt{d-1}$.}
\label{alfa}
\end{figure}

For a given $d >1$ we have studied the universal behavior 
as a function of $N$, the size of our system. We assume the 
blob (and only the blob, not the stalk)
is characterized by a Hausdorff dimension $d_H$, measuring 
the space-time volume enclosed inside a geodesic\footnote{We use 
here as geodesic distance the link distance between two 
vertices on the $\phi^3$ graph.} ball of 
radius $r$: $V(r)\propto r^{d_H}$. We then expect that the quantity 
\begin{equation}\label{rho}
\rho (\tau) = V(t) N^{1/d_H -1},
\end{equation}
plotted as a function of the scaled time variable $\tau = t/N^{1/d_H}$ 
should be a universal function, independent of the total volume $N$.
Fig. \ref{scaling} illustrates this scaling for $d = 6, 4, 3$ and $2$ 
with $d_H = 3$. The continuous curve is a plot of
\begin{equation}\label{rho0}
\rho_0(\tau) =  \frac{2\alpha}{\pi} \cos^2(\alpha\tau)
\end{equation}
with a coefficient $\alpha$ depending on $d$.
The dependence of $1/\alpha(d)$ is presented in the Fig.\ \ref{alfa}, 
from which we find that $\alpha$ increases with $d$.

\subsection{The effective action}

It is an  obvious question what kind of action 
creates the observed blob-structure? From the studies 
of 4d CDT we know how to extract the effective action \cite{ajl4d, planck}
for such a system. We want to get the effective action as a function 
of $V(t)$, and  we can obtain the quadratic part by analyzing 
the covariance matrix 
\begin{equation}
C_{t,t'} = \langle V(t) V(t')\rangle - \langle V(t)\rangle\langle V(t')\rangle
\end{equation}
for a given $d$ and $N$. As argued in \cite{ajl4d,planck}, 
inverting such a matrix gives an "inverse propagator" 
$P(t,t')$, or a matrix of second derivatives
of $S_{eff}[V(t)]$ with respect to $V(t)$ and $V(t')$. 
We find the structure very similar to that in the 4D case, with non-zero 
matrix elements  concentrated on the  diagonal and the two off-diagonals. 
The observed behavior is consistent with the effective action
\begin{equation}
S_{eff} = \frac{1}{\Gamma} \sum_t 
\left( \frac{(V(t)-V(t+1))^2}{V(t)+V(t+1)} -\lambda V(t)\right).
\label{Seff}
\end{equation}
In this equation $\lambda$ should be viewed as a Lagrange multiplier
which is determined by the condition that the integral of $V(t)$ 
is equal to $\bar{N}$, see eq.\ (\ref{N}). The equation represents 
a naive discretization of the mini-superspace action in three dimensions,
in accordance with our measurement that $d_H =3$. 

\begin{figure}[t]
\centerline{\scalebox{1.0}{\includegraphics{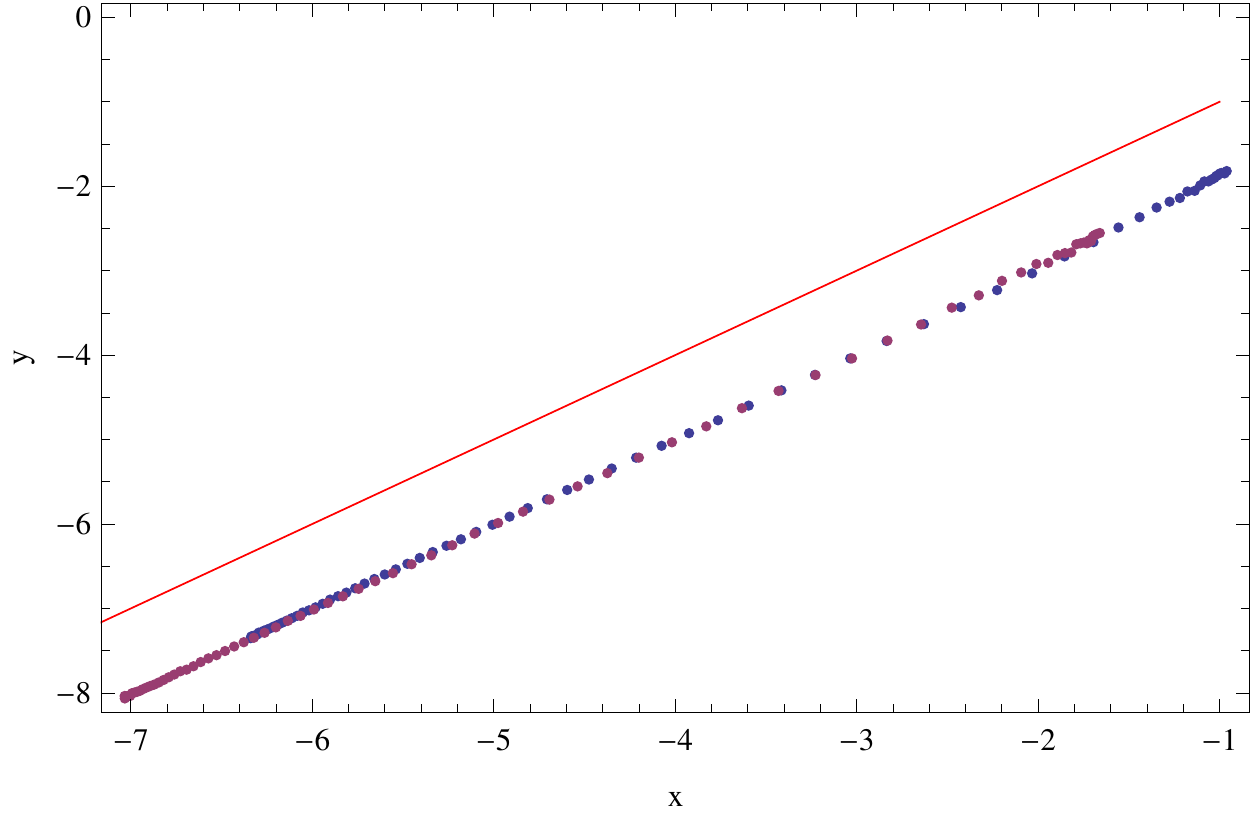}}}
\caption{ The diagonal and the first off-diagonal of the inverse matrix 
versus the theoretical values (see text).}
\label{covariance}
\end{figure}

The figure \ref{covariance} shows, on a log-log plot, the comparison of
the measured values of the diagonal and the first off-diagonal matrix 
elements (multiplied by a factor (-2))  (the $y$ axis) versus the same values 
calculated as derivatives of (\ref{Seff}), using experimental values 
of $\langle V(t)\rangle$ (the $x$ axis). The system is for  
$d=4$, $N = 128000$ and $T= 300$. The straight line shown on the plot is 
$y = x$. The (constant) shift between the experimental plots and 
the straight line can be used to determine the value of $\Gamma$.

\begin{figure}[t]
{\scalebox{0.7}{\includegraphics{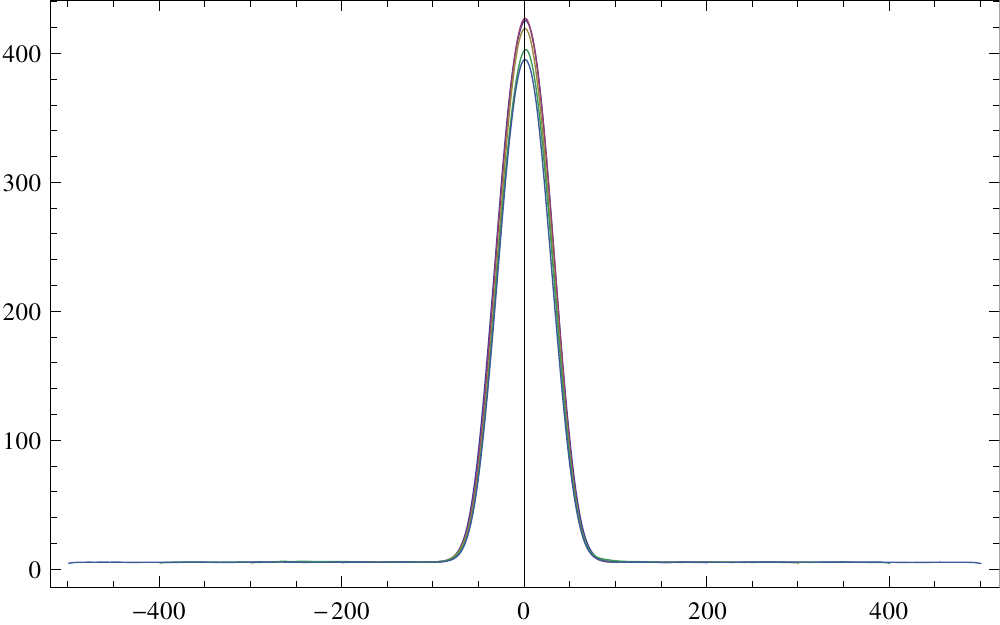}}}
{\scalebox{0.7}{\includegraphics{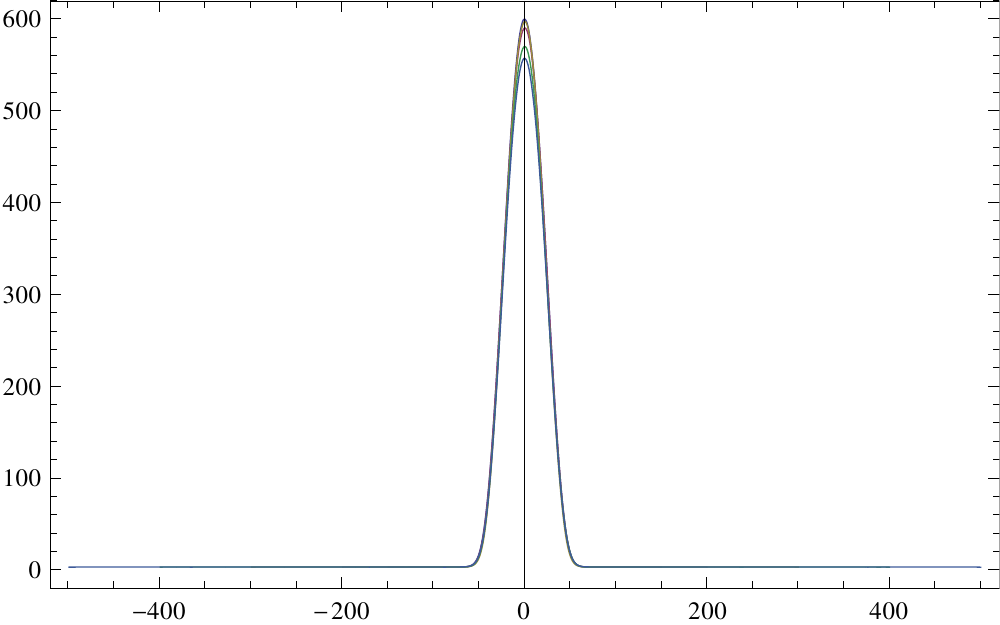}}}
\caption{The dependence of the semi-classical volume distribution on $T$ 
for two and three scalar fields, volume 64k. In both cases 
we use $T= 200, ~240, ~320, ~400, ~480, ~600$.}
\label{fixedvolume2}
\end{figure}

\begin{figure}[th!]
{\scalebox{0.7}{\includegraphics{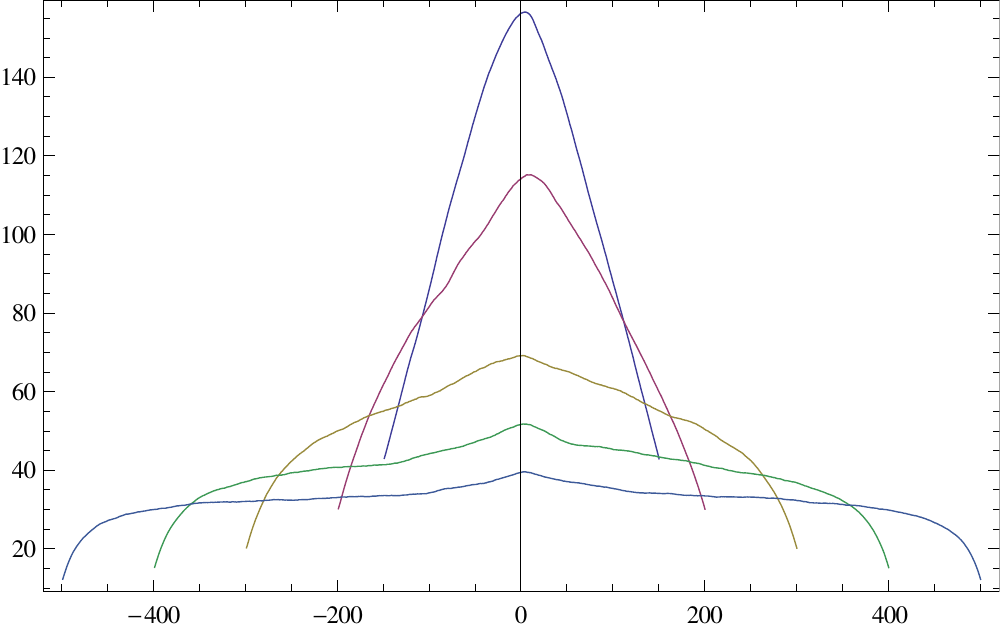}}}
{\scalebox{0.7}{\includegraphics{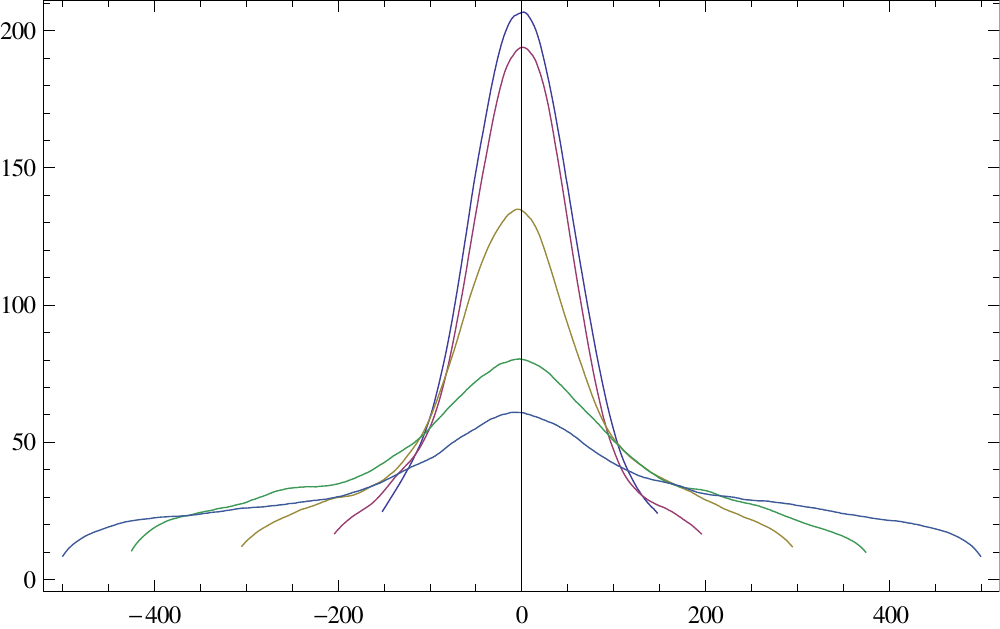}}}
\caption{The dependence of the semi-classical volume distribution on 
$T$ for pure gravity and one scalar field, volume 64k. 
In both cases we use $T = 200, ~240, ~320, ~400, ~480, ~600$.}
\label{fixedvolume1}
\end{figure}

Applying the same method  to extract the semi-classical  volume 
distributions for $d=0$ and $d=1$ where we have 
``unbroken symmetry'', i.e. no  blob-formation,  leads to
configurations centered around the largest volume. 
As a  consequence the average volume distribution is not
flat (as we would expect for the toroidal geometry with no blob-formation), 
but has a maximum at zero time by construction.  
More specifically, since there is no dominant blob, but 
rather a a sequence of spatial volume fluctuations 
with varying length in time $t$,
we pick for each configuration 
the largest fluctuation and measure distributions 
centered around this fluctuation. Clearly, the average 
distributions $\langle V(t) \rangle$ obtained this way for $d=0$ and $d=1$
are very different from the distributions observed for $d>1$, having 
no stalk, but spreading out when $T$ is increased and $N$ kept fixed
(see Fig.\ \ref{fixedvolume1} and compare to Fig.\ \ref{fixedvolume2}
where $d>1$ and we have blob-formation).

\vspace{6pt}

\begin{figure}[th!]
\centerline{\scalebox{0.8}{\includegraphics{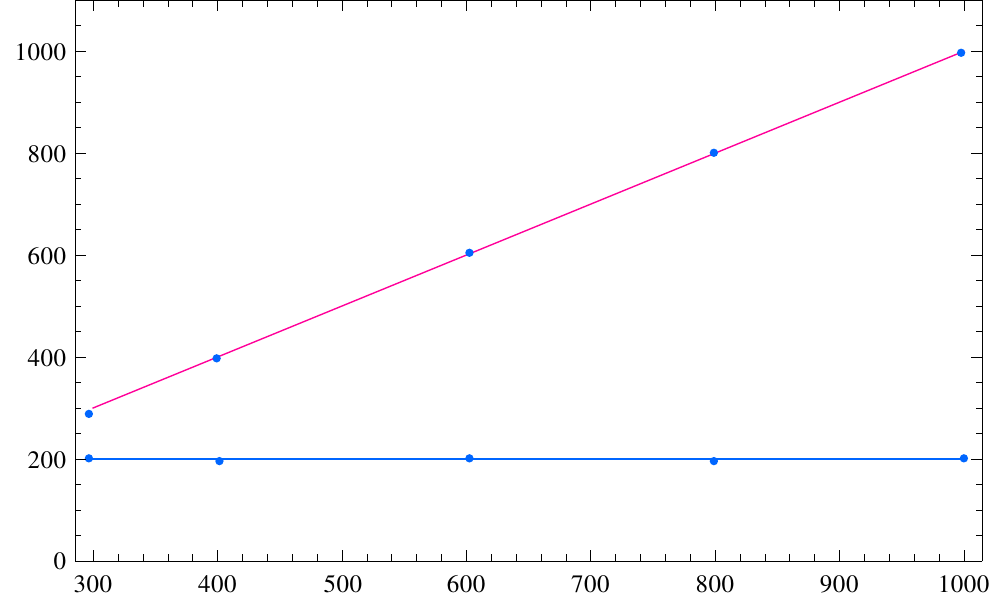}}}
\caption{The relation between $w_T(d)$ on $d$ with different scalar 
fields in fixed volume 64k. The red curve is pure gravity  and 
the blue curve is $d = 2$. We use $T = 300, ~400,~ 600, ~800, ~1000$ 
for both cases. }
\label{width} 
\end{figure}

To summarize:

\vspace{-8pt}

 \begin{itemize}
\item For $d>1$ the shape of the semi-classical distribution is 
practically independent of $T$ (see the Fig \ref{fixedvolume2}). 
There is a small shift near the maximum, since in all plots we have 
the same volume. 
\item For $d \leq 1$ the width $w_T(d)$  grows linearly with $T$ when 
the period $T$ is increased and this effect is shown on the  
Fig. \ref{fixedvolume1} for $d=0$ (pure gravity) and for $d = 1$.
The  dependence of the width $w_T(d)$ on $d$ and $T$ is shown in 
Fig.\  \ref{width}.
\end{itemize}

\section{Discussion and conclusion}\label{Conclusions}

We have studied the causal dynamical triangulation (CDT) 
model for quantum gravity in two dimensions 
coupled to $d$  massless scalar fields. The topology
of the space-time was chosen to be $S^1\times S^1$,
i.e.\ both space and (Euclidean) time were chosen periodic.  
 
Knowing that pure  two-dimensional CDT can be viewed as 
an effective theory of pure ($c=0$) Euclidean quantum gravity where 
baby universes have been integrated out, we asked
if one can observe any trace of the $c=1$ barrier known
in Euclidean quantum gravity 
if we coupled conformal matter to CDT 2d gravity.
Somewhat surprisingly the result is not only yes, but the numerical 
signal is much clearer than what is seen in the (lattice versions) of 
Euclidean quantum gravity \cite{adjt,davidBP}. 

When $d >1$ we observe what we have called 
``semiclassical'' configurations which
dominate the path integral. These are ``universes'' which have
a spatial volume distribution $V(t)$ described by (\ref{rho0}),
i.e. a distribution  much like one would have 
if the blob had the geometry of $S^3$, or an elongated version
of $S^3$ as described in \cite{deformed}.  
A corresponding mini-superspace action is given by (\ref{Seff}).
From the finite sized scaling relation (\ref{rho}) 
which is very well satisfied we obtain the Hausdorff 
dimension of the  blob to be 3. This seems to be valid for 
all $d>1$, but with the blob structure more pronounced with 
increasing $d$. It is also consistent with the results 
obtained in \cite{aal}, where the system of 8 Ising spins coupled to 
2d CDT was studied. Also here a blob with $d_H=3$ was observed
(although it was not checked if the distribution of $V(t)$ 
was given by (\ref{rho0})). 
Since this is a very different matter system, seemingly both the 
formation of a blob and its Hausdorff dimension 
are independent of the kind of conformal matter coupled to the CDT 
geometry, as long as $c>1$.  

Since the effective action by construction reproduces
the volume-volume correlator, this action will also describe
(to quadratic approximation) the quantum fluctuations around
the semiclassical blob.     
 
The way we use the $d$ Gaussian fields in the Monte Carlo simulations
forces $d$ to be a non-negative integer. Thus it is difficult to 
address the exact nature of the CDT $c=1$ ``barrier'' since we start at 
$c=2$. However, since the action is Gaussian it is in 
principle possible to integrate out the matter fields for 
each triangulation ${\cal T}$. Let $D_{\cal T}(i,j)$ denote the 
so-called coincidence matrix of the triangulation ${\cal T}$, an $N\times N$
matrix which is 3 in the diagonal and -1 if $i$ and $j$ are neighboring 
vertices in the $\phi^3$-graph ${\cal T}$. Let $\Delta'({\cal T})$
be the determinant of $D_{\cal T}(i,j)$ with the zero mode removed.
We have 
\begin{equation}\label{determinant}
\int \prod_{i,\mu} d'x_i^\mu~  e^{-S_{Gauss}({\cal T},x^\mu)} \propto
\Big(\Delta'({\cal T})\Big)^{-d/2},
\end{equation}     
where $d'$ symbolizes that the integration excludes 
the translational zero-mode. Thus the partition function ${\cal Z}$
from eq.\ (\ref{pathintegral}) can be written as 
\begin{equation}\label{path1}
{\cal Z} = \sum_{{\cal T}} \frac{1}{C_T} \; 
e^{-\Lambda N({\cal T}-\epsilon(N({\cal T})-\bar{N})^2)}  
\Big(\Delta'({\cal T})\Big)^{-d/2}.
\end{equation}  
The advantage of this representation is that it is a continuous function
of the number of Gaussian fields and we can thus use it to approach 
$d=1$ from above. However, the disadvantage of the expression is 
that we have to calculate the $N\times N$ determinant when 
we update geometries. It can be done when $N$ is less than a few 
hundred \cite{determinant}, but such $N$ might be too small to 
signal clearly a phase transition.

The geometric structure observed here  is similar to that 
observed in CDT formulation of the 3D and 4D quantum gravity. 
Also in this case the geometries which dominate
the path integral are {\it not} (Euclidean) time-translational 
invariant configurations,  
and the systems in the so-called  (Euclidean) de Sitter phase can be viewed
as semi-classical 3D and  4D spheres with superimposed quantum fluctuations. 
Hopefully the fact that matter seems to trigger the 
same kind of configurations will help us 
understand in more detail the mechanism responsible for  
creating the regular spherical semi-classical geometry in higher 
dimensional CDT.

Clearly our ``spheres'' with Hausdorff dimension 3 cannot 
be three-dimensional spheres in an ordinary sense since
they are constructed by triangles, i.e.\ two-dimensional building
blocks. A measurement of the spectral dimension
of the spheres would be very interesting, in particular since 
the spectral dimension in higher dimensional CDT seemingly 
has quite surprising characteristics linking it both to 
Ho\v rava-Lifshitz gravity and UV fix points in 
asymptotic safety scenarios \cite{safety}. 

Another interesting aspect to investigate is the effect of 
a mass term. If we add a mass  to the Gaussian field, the
field theory is no longer a conformal field theory. Naively, 
since we try to formulate a quantum theory of gravity,  one would 
expect such a mass term to interact strongly with the geometry.
However, the main effect in 2d may just be that the fields are 
forced to lie in the neighborhood of zero field 
and as a  consequence the field part of the action becomes irrelevant. 
Thus we can  maybe expect a phase transition back to the 
$c \leq 1$ phase as a function of the mass. It might
give us an unexpected easy way to study the 
phase transition related to the $c=1$ barrier.

\vspace{1cm}

\noindent {\bf Acknowledgments.} Zhang thanks  
J. Gizbert-Studnicki for help. This work is partly supported by 
the International PhD Projects Programme of the Foundation for Polish 
Science within the European Regional Development Fund of the European 
Union, agreement no. MPD/2009/6. 
JA thanks the Institute of 
Theoretical Physics and the Department of Physics and Astronomy 
at Utrecht University for hospitality and financial support. 
J.A. and A.G. acknowledge financial 
support by  the Danish Research Council (FNU) from the grant 
``quantum gravity and the role of black holes''.


\vspace{1cm}

\begin{thebibliography}{99}

\bibitem{old}
  F.~David,
 {\it Planar Diagrams, Two-Dimensional Lattice Gravity and Surface Models,}
  Nucl.\ Phys.\  {\bf B257 } (1985)  45.\\
A.~Billoire and F.~David,
{\it Microcanonical simulations of randomly triangulated planar random
surfaces},
Phys.\ Lett.\  B\ {\bf 168} (1986) 279-283.\\
  V.~A.~Kazakov, A.~A.~Migdal, I.~K.~Kostov,
 {\it Critical Properties of Randomly Triangulated Planar Random Surfaces,}
  Phys.\ Lett.\  {\bf B157 } (1985)  295-300.\\
D.V.~Boulatov, V.A.~Kazakov, I.K.~Kostov and A.A.~Migdal,
{\it Analytical and numerical study of the model of dynamically triangulated
random surfaces},
Nucl.\ Phys.\  B\ {\bf 275} (1986) 641-686.\\
J.~Ambjorn, B.~Durhuus and J.~Fr\"ohlich,
{\it Diseases of triangulated random surface models, and possible cures},
Nucl.\ Phys.\  B\ {\bf 257} (1985) 433-449;\\
J.~Ambjorn, B.~Durhuus, J.~Fr\"ohlich and P.~Orland,
{\it The appearance of critical dimensions in regulated string theories},
Nucl.\ Phys.\  B\ {\bf 270} (1986) 457-482.\\
  J.~Jurkiewicz, A.~Krzywicki and B.~Petersson,
{\it A Numerical Study Of Discrete Euclidean Polyakov Surfaces,}
  Phys.\ Lett.\ B {\bf 168} (1986) 273.\\
  J.~Jurkiewicz, A.~Krzywicki and B.~Petersson,
 {\it A Grand Canonical Ensemble Of Randomly Triangulated Surfaces,}
  Phys.\ Lett.\ B {\bf 177} (1986) 89.



\bibitem{polyakov}
  A.~M.~Polyakov,
 {\it Quantum Geometry of Bosonic Strings,}
  Phys.\ Lett.\ B {\bf 103} (1981) 207.

\bibitem{old1}
J.~Ambjorn and J.~Jurkiewicz,
{\it Four-dimensional simplicial quantum gravity},
Phys.\ Lett.\  B\ {\bf 278} (1992) 42-50.\\
  
  
  
  
M.E.~Agishtein and A.A.~Migdal,
{\it Simulations of four-dimensional simplicial quantum gravity},
Mod.\ Phys.\ Lett.\  A\ {\bf 7} (1992) 1039-1062.\\
  J.~Ambjorn, S.~Varsted,
{\it Three-dimensional simplicial quantum gravity,}
  Nucl.\ Phys.\  {\bf B373 } (1992)  557-580.\\
J.~Ambjorn, S.~Varsted,
{\it Entropy estimate in three-dimensional simplicial quantum gravity,}
  Phys.\ Lett.\  {\bf B266 } (1991)  285-290.\\
  J.~Ambjorn, D.~V.~Boulatov, A.~Krzywicki, S.~Varsted,
{\it The Vacuum in three-dimensional simplicial quantum gravity,}
  Phys.\ Lett.\  {\bf B276 } (1992)  432-436.\\
  M.~E.~Agishtein, A.~A.~Migdal,
 {\it Three-dimensional quantum gravity as dynamical triangulation,}
  Mod.\ Phys.\ Lett.\  {\bf A6 } (1991)  1863-1884.\\
  D.~V.~Boulatov, A.~Krzywicki,
 {\it On the phase diagram of three-dimensional simplicial quantum gravity,}
  Mod.\ Phys.\ Lett.\  {\bf A6 } (1991)  3005-3014.



\bibitem{firstorder}
  P.~Bialas, Z.~Burda, A.~Krzywicki, B.~Petersson,
 {\it Focusing on the fixed point of 4-D simplicial gravity,}
  Nucl.\ Phys.\  {\bf B472 } (1996)  293-308.
  [hep-lat/9601024].\\
  S.~Catterall, R.~Renken, J.~B.~Kogut,
 {\it Singular structure in 4-D simplicial gravity,}
  Phys.\ Lett.\  {\bf B416 } (1998)  274-280.
  [hep-lat/9709007].

\bibitem{crinkled}
  S.~Bilke, Z.~Burda, A.~Krzywicki, B.~Petersson, 
J.~Tabaczek and G.~Thorleifsson,
{\it 4-D simplicial quantum gravity interacting with gauge matter fields,}
  Phys.\ Lett.\ B {\bf 418} (1998) 266;
  [hep-lat/9710077].\\
{\it 4-D simplicial quantum gravity: 
Matter fields and the corresponding effective action,}
  Phys.\ Lett.\ B {\bf 432} (1998) 279
  [hep-lat/9804011].\\
  J.~Ambjorn, K.~N.~Anagnostopoulos and J.~Jurkiewicz,
{\it Abelian gauge fields coupled to simplicial quantum gravity,}
  JHEP {\bf 9908} (1999) 016
  [hep-lat/9907027].\\
  S.~Horata, H.~S.~Egawa, N.~Tsuda, T.~Yukawa,
 {\it Phase structure of four-dimensional 
simplicial quantum gravity with a U(1) gauge field,}
  Prog.\ Theor.\ Phys.\  {\bf 106 } (2001)  1037-1050.
  [hep-lat/0004021].



\bibitem{al1}
J. Ambjorn and R. Loll, 
{\it Nonperturbative Lorentzian quantum gravity, 
causality and topology change,} 
Nucl.Phys.B {\bf 536} (1998) 407-434, [hep-th/9805108].


\bibitem{ajl3d}
J.~Ambjorn, J.~Jurkiewicz and R.~Loll:
{\it Non-perturbative 3d Lorentzian quantum gravity,}
Phys.\ Rev.\  D {\bf 64} (2001) 044011 [hep-th/0011276];


\bibitem{ajl4d}
J. Ambjorn, J. Jurkiewicz and R. Loll, 
{\it Dynamically triangulating Lorentzian quantum gravity,} 
Nucl.Phys.B {\bf 610} (2001), 347-382 [hep-th/0105267].\\
{\it A non-perturbative Lorentzian path integral for gravity,}
Phys.\ Rev.\ Lett.\  {\bf 85} (2000) 924 [hep-th/0002050].



\bibitem{problems}
  S.~Catterall, G.~Thorleifsson, J.~B.~Kogut, R.~Renken,
 {\it Singular vertices and the triangulation space of the D sphere,}
  Nucl.\ Phys.\  {\bf B468 } (1996)  263-276.
  [hep-lat/9512012].\\
P.~Bialas, Z.~Burda, B.~Petersson and J.~Tabaczek,
{\it Appearance of mother universe and singular vertices in random
geometries,}
Nucl.\ Phys.\ B\ 495 (1997) 463-476 [hep-lat/9608030].



\bibitem{horava1}
  P.~Horava,
 {\it Quantum Gravity at a Lifshitz Point,}
  Phys.\ Rev.\  {\bf D79 } (2009)  084008.
  [arXiv:0901.3775 [hep-th]].\\
  P.~Horava, C.~M.~Melby-Thompson,
 {\it General Covariance in Quantum Gravity at a Lifshitz Point,}
  Phys.\ Rev.\  {\bf D82 } (2010)  064027.
  [arXiv:1007.2410 [hep-th]].\\
 {\it Anisotropic Conformal Infinity,}
  [arXiv:0909.3841 [hep-th]].


\bibitem{spectral1}
J.~Ambjorn, J.~Jurkiewicz and R.~Loll,
{\it Spectral dimension of the universe}, 
Phys.\ Rev.\ Lett.\ {\bf 95} (2005) 171301
[hep-th/0505113].


\bibitem{spectral2}
P.~Ho\v rava:
{\it Spectral dimension of the universe in quantum 
gravity at a Lifshitz point,}
Phys.\ Rev.\ Lett.\  {\bf 102} (2009) 161301 [arXiv:0902.3657, hep-th].


\bibitem{al-samo}
J.~Ambjorn, A.~G\"orlich, S.~Jordan, J.~Jurkiewicz and R.~Loll,
{\it CDT meets Horava-Lifshitz gravity},
Phys.\ Lett.\ B {\bf 690} (2010) 413-419 [arXiv:1002.3298, hep-th].





\bibitem{newD3}
  C.~Anderson, S.~Carlip, J.~H.~Cooperman, P.~Horava, 
R.~Kommu and P.~R.~Zulkowski,
{\it Quantizing Horava-Lifshitz Gravity via Causal Dynamical Triangulations,}
  arXiv:1111.6634 [hep-th].




\bibitem{semiclassical}
J.~Ambjorn, J.~Jurkiewicz and R.~Loll,
{\it Emergence of a 4D world from causal quantum gravity},
Phys.\ Rev.\ Lett.\ {\bf 93} (2004) 131301 [hep-th/0404156].\\
{\it Reconstructing the universe},
Phys.\ Rev.\ D\ {\bf 72} (2005) 064014  [hep-th/0505154].

\bibitem{planck}
J.~Ambjorn, A.~G\"orlich, J.~Jurkiewicz and R.~Loll,
{\it Planckian birth of the quantum de Sitter universe},
Phys.\ Rev.\ Lett.\ {\bf 100} (2008) 091304 [arXiv:0712.2485, hep-th].\\
{\it The nonperturbative quantum de Sitter universe},
Phys.\ Rev.\  D {\bf 78} (2008) 063544 [arXiv:0807.4481, hep-th].\\
{\it Semiclassical universe from first principles},
Phys.\ Lett.\ B {\bf 607} (2005) 205-213
[hep-th/0411152].\\
J.~Ambjorn, A.~G\"orlich, J.~Jurkiewicz and R.~Loll,
{\it Geometry of the quantum universe},
Phys.\ Lett.\ B {\bf 690} (2010) 420-426 [arXiv:1001.4581, hep-th].


\bibitem{secondorder}
  J.~Ambjorn, S.~Jordan, J.~Jurkiewicz and R.~Loll,
 {\it A Second-order phase transition in CDT,}
  Phys.\ Rev.\ Lett.\  {\bf 107} (2011) 211303
  [arXiv:1108.3932 [hep-th]].



\bibitem{3dCDT}
J.\ Ambjorn, J.\ Jurkiewicz, R.\ Loll and G.\ Vernizzi:
{\it Lorentzian 3d gravity with wormholes via matrix models},
JHEP {\bf 0109} (2001) 022 [hep-th/0106082];\\
{\it 3D Lorentzian quantum gravity from the asymmetric ABAB matrix model},
Acta Phys.\ Polon.\ B\ {\bf 34} (2003) 4667-4688 [hep-th/0311072];\\
J.~Ambjorn, J.~Jurkiewicz and R.~Loll:
{\it Renormalization of 3d quantum gravity from matrix models,}
Phys.\ Lett.\ B\ {\bf 581} (2004) 255-262 [hep-th/0307263].\\
D.~Benedetti, R.~Loll and F.~Zamponi:
{\it (2+1)-dimensional quantum gravity as the continuum limit of causal
dynamical triangulations}, 
Phys.\ Rev.\ D\ {\bf 76} (2007) 104022 [arXiv:0704.3214, hep-th].


\bibitem{laiho}
  J.~Laiho and D.~Coumbe,
 {\it Evidence for Asymptotic Safety from Lattice Quantum Gravity,}
  Phys.\ Rev.\ Lett.\  {\bf 107} (2011) 161301
  [arXiv:1104.5505 [hep-lat]].

\bibitem{ack}
  J.~Ambjorn, J.~Correia, C.~Kristjansen, R.~Loll,
{\it On the relation between Euclidean and Lorentzian 2-D quantum gravity,}
  Phys.\ Lett.\ B  {\bf 475 } (2000)  24-32.
  [hep-th/9912267].


\bibitem{generalizedCDT}
J.~Ambjorn, R.~Loll, W.~Westra and S.~Zohren,
{\it Putting a cap on causality violations in CDT,}
JHEP {\bf 0712} (2007) 017 [arXiv:0709.2784, gr-qc];\\
{\it A matrix model for 2D quantum gravity defined by causal dynamical
triangulations},
Phys.\ Lett.\  B\ {\bf 665} (2008) 252-256 [arXiv:0804.0252, hep-th];\\
{\it A new continuum limit of matrix models,}
Phys.\ Lett.\  B {\bf 670} (2008) 224 [arXiv:0810.2408, hep-th].\\
J.~Ambjorn, R.~Loll, Y.~Watabiki, W.~Westra and S.~Zohren:
{\it A string field theory based on causal dynamical triangulations,}
JHEP {\bf 0805} (2008) 032 [arXiv:0802.0719, hep-th].




\bibitem{transfer}
  H.~Kawai, N.~Kawamoto, T.~Mogami and Y.~Watabiki,
 {\it Transfer matrix formalism for two-dimensional quantum 
gravity and fractal structures of space-time,}
  Phys.\ Lett.\ B {\bf 306} (1993) 19
  [hep-th/9302133].



\bibitem{aw}
J.~Ambjorn and Y.~Watabiki,
{\it Scaling in quantum gravity,}
Nucl.\ Phys.\  B {\bf 445} (1995) 129 [hep-th/9501049].



\bibitem{ajw}
J.~Ambjorn, J.~Jurkiewicz and Y.~Watabiki,
{\it On the fractal structure of two-dimensional quantum gravity,}
Nucl.\ Phys.\ B\ {\bf 454} (1995) 313-342 [hep-lat/9507014].



\bibitem{bowick}
  S.~Catterall, G.~Thorleifsson, M.~J.~Bowick, V.~John,
 {\it Scaling and the fractal geometry of two-dimensional quantum gravity,}
  Phys.\ Lett.\  {\bf B354 } (1995)  58-68.
  [hep-lat/9504009].


\bibitem{nishimura}
  H.~Aoki, H.~Kawai, J.~Nishimura and A.~Tsuchiya,
 {\it Operator product expansion in two-dimensional quantum gravity,}
  Nucl.\ Phys.\ B {\bf 474} (1996) 512
  [hep-th/9511117].


\bibitem{KPZ}
  V.~G.~Knizhnik, A.~M.~Polyakov and A.~B.~Zamolodchikov,
{\it Fractal Structure of 2D Quantum Gravity,}
  Mod.\ Phys.\ Lett.\ A {\bf 3} (1988) 819.\\
  F.~David,
{\it Conformal Field Theories Coupled to 2D Gravity in the Conformal Gauge,}
  Mod.\ Phys.\ Lett.\ A {\bf 3} (1988) 1651.\\
  J.~Distler and H.~Kawai,
{\it Conformal Field Theory and 2D Quantum Gravity Or 
Who's Afraid of Joseph Liouville?,}
  Nucl.\ Phys.\ B {\bf 321} (1989) 509.


\bibitem{durhuus}
  B.~Durhuus, J.~Frohlich and T.~Jonsson,
 {\it Critical Behavior in a Model of Planar Random Surfaces,}
  Nucl.\ Phys.\ B {\bf 240} (1984) 453;
  Phys.\ Lett.\ B  {\bf 137} (1984) 93.



\bibitem{ad}
  J.~Ambjorn and B.~Durhuus,
 {\it Regularized Bosonic Strings Need Extrinsic Curvature,}
  Phys.\ Lett.\ B {\bf 188} (1987) 253.



\bibitem{aal}
J.~Ambjorn, K.N.~Anagnostopoulos and R.~Loll:
{\it A new perspective on matter coupling in 2d quantum gravity},
Phys.\ Rev.\  D\ {\bf 60} (1999) 104035 [hep-th/9904012].

\bibitem{aalp}
J.~Ambjorn, K.N.~Anagnostopoulos, R.~Loll and I.~Pushkina:
{\it Shaken, but not stirred - Potts model coupled to quantum gravity,}
Nucl.\ Phys.\  B {\bf 807} (2009) 251 [arXiv:0806.3506, hep-lat].


\bibitem{jain}
  S.~Jain and S.~D.~Mathur,
{\it World sheet geometry and baby universes in 2-D quantum gravity,}
  Phys.\ Lett.\ B {\bf 286} (1992) 239
  [hep-th/9204017].



\bibitem{ajt}
  J.~Ambjorn, S.~Jain, G.~Thorleifsson,
 {\it Baby universes in 2-d quantum gravity,}
  Phys.\ Lett.\  {\bf B307 } (1993)  34-39.
  [hep-th/9303149].

\bibitem{adj}
  J.~Ambjorn, B.~Durhuus and T.~Jonsson,
 {\it A Solvable 2-d gravity model with $\gamma  > 0$,}
  Mod.\ Phys.\ Lett.\ A {\bf 9} (1994) 1221
  [hep-th/9401137].


\bibitem{spikes}
M.E. Cates,
Europhysics Lett. B {\bf 719} (1988).\\
  A.~Krzywicki,
 {\it On The Stability Of Random Surfaces,}
  Phys.\ Rev.\ D {\bf 41} (1990) 3086.



\bibitem{charlotte}
  P.~Di Francesco, E.~Guitter and C.~Kristjansen,
{\it Integrable 2-D Lorentzian gravity and random walks,}
  Nucl.\ Phys.\ B {\bf 567} (2000) 515
  [hep-th/9907084].




\bibitem{barkama}
M.E.J Newman and T. Barkama, 
{\it Monte Carlo Methods in Statistical Physics},
Oxford University Press 2002.



\bibitem{adjt}
  J.~Ambjorn, B.~Durhuus, T.~Jonsson and G.~Thorleifsson,
{\it Matter fields with $c > 1$ coupled to 2-d gravity,}
  Nucl.\ Phys.\ B {\bf 398} (1993) 568
  [hep-th/9208030].


\bibitem{davidBP}
  F.~David,
 {\it A Scenario for the $c > 1$ barrier in noncritical bosonic strings,}
  Nucl.\ Phys.\ B {\bf 487} (1997) 633
  [hep-th/9610037].

\bibitem{deformed}
  D.~Benedetti and J.~Henson,
 {\it Spectral geometry as a probe of quantum spacetime,}
  Phys.\ Rev.\ D {\bf 80} (2009) 124036
  [arXiv:0911.0401 [hep-th]].



\bibitem{determinant}
  J.~Ambjorn, P.~de Forcrand, F.~Koukiou and D.~Petritis,
 {\it Monte Carlo Simulations Of Regularized Bosonic Strings,}
  Phys.\ Lett.\ B {\bf 197} (1987) 548.

\bibitem{safety}
O.~Lauscher and M.~Reuter,
{\it Fractal spacetime structure in asymptotically safe gravity,}
JHEP {\bf 0510} (2005) 050 [hep-th/0508202].







\end{thebibliography}
\end{document}